\documentclass[sigconf]{acmart}

\usepackage{enumitem}
\usepackage{xcolor}
\usepackage[most]{tcolorbox}
\usepackage{adjustbox}
\usepackage{wrapfig}
\usepackage{graphicx}
\usepackage{subcaption}
\usepackage{tikz}
\usepackage{booktabs}

\definecolor{shadecolor}{rgb}{0.94, 0.97, 1.0}
\definecolor{desertsand}{rgb}{0.93, 0.79, 0.69}
\definecolor{bubbles}{rgb}{0.91, 1.0, 1.0}
\definecolor{bubblegum}{rgb}{0.99, 0.76, 0.8}

\newcommand{\eg}{\textit{e.g.,}}
\newcommand{\ie}{\textit{i.e.,}}
\newcommand{\etal}{\textit{et al.}}

\AtBeginDocument{%
  \providecommand\BibTeX{{%
    \normalfont B\kern-0.5em{\scshape i\kern-0.25em b}\kern-0.8em\TeX}}}

\setcopyright{acmcopyright}
\copyrightyear{2018}
\acmYear{2018}
\acmDOI{XXXXXXX.XXXXXXX}

\acmPrice{15.00}
\acmISBN{978-1-4503-XXXX-X/18/06}

\begin{document}

\title[Developers' Trust Perceptions through Urgency and Reputation]{An Eye for Trust: An Exploration of Developers' Trust Perceptions Through Urgency and Reputation}



\author{Sara Yabesi}
\affiliation{%
  \institution{Polytechnique Montr\'eal}
  \city{Montr\'eal}
  \country{Canada}}
\email{sara.yabesi@polymtl.ca}

\author{Mahta Amini}
\affiliation{%
  \institution{Polytechnique Montr\'eal}
  \city{Montr\'eal}
  \country{Canada}}
\email{mahta.amini@polymtl.ca}

\author{Jelena Ristic}
\affiliation{%
  \institution{McGill University}
  \city{Montr\'eal}
  \country{Canada}}
\email{jelena.ristic@mcgill.ca}

\author{Zohreh Sharafi}
\affiliation{%
  \institution{Polytechnique Montr\'eal}
  \city{Montr\'eal}
  \country{Canada}}
\email{zohreh.sharafi@polymtl.ca}

\renewcommand{\shortauthors}{Yabesi, et al.}

\begin{abstract}

Code reuse is a widespread practice across software development projects, suggesting an inherent trust in the reused code. Yet, there is a lack of a fundamental understanding of developers' trust and how various factors mold their trust--based cognitive processes. 
Drawing from the psychology of compliance and trust, we
present the results of the first controlled experiment (n=37) which uses  eye tracking to explore how urgency (represented by code priority level) and reputation (represented by the experience level of the code's author) influence developers' perceptions of code trustworthiness. 

Our research revealed that the priority assigned to a code patch significantly influenced developers' code review behavior, impacting their evaluation time, cognitive load, and perceived quality. However, the decision to incorporate and implement the code was not affected . 
Eye tracking data revealed that there were variations in overall visual code scanning and the distribution of attention across identical code patches labeled as written by senior vs. junior developers. Yet, there were no significant performance differences. Moreover, our participants nominate code functionality, quality, and comprehensibility as primary factors in code evaluation. 

Despite noticeable changes in code review behavior, our participants surprisingly overlooked the substantial influence of urgency and reputation on their decisions to review and reuse code changes. This study takes the next step toward a better understanding of trust in software engineering and may inform future research about code review platforms and guidelines, code reuse, and automated code generation.

\end{abstract}




\received{20 February 2007}
\received[revised]{12 March 2009}
\received[accepted]{5 June 2009}

\maketitle

\section{Introduction}

In The Mythical Man Month (1974), Fred Brooks suggests that reusable code costs three times as much to develop as single--use code. However, code reuse is prevalent across various software projects~\cite{p-Mocku-reuse-icse-2007,p-haefliger-reuse-code-2008,p-alarcon-heuristic-systematic-2017}. The profound impact of code reuse becomes undeniably apparent when analyzing its scale in relation to widely--recognized open--source software (OSS) projects, such as log4j and jUnit. 
The code reuse of such software saved roughly 316 thousand person-years and represents tens of billions of dollars in development expenditures~\cite{p-Ampatzoglou-reuse-OSS-2013}.

Reuse implies that the artifact is trustworthy, as the developer primarily aims to use it in a system for which it was not initially intended. Still, there is a limited understanding of why developers have trust in software or artifacts they have not developed. This paper focuses on code reuse from a human trust perspective.

During a cognitive task analysis, the trustworthiness of an artifact is influenced by various interrelated factors~\cite{alarcon-hsm-computer-code-trust-2018,lu-sarter-eye-trust}, such as urgency and trust cues~\cite{p-naidoo-phishing-trust-2015}. Urgency can alter how individuals process information and make decisions, amplifying their reliance on  trust cues such as the source reputation. Trust cues, in turn, can significantly affect how individuals respond to urgent situations~\cite{p-naidoo-phishing-trust-2015}. Leveraging principles from the psychology of trust, this paper performs an unprecedented examination to study how urgency and reputation impact developers’ trust perception.

In the field of psychology, various metrics have been used to deduce and quantify trust. Past studies concentrated on self-reporting methods, \eg{} think-aloud protocols, surveys, and interviews to evaluate trust~\cite{lu-sarter-eye-trust}. However, these methods are prone to the Hawthorne effect, where the presence of an observer can influence the behavior of participants ~\cite{p-etelapelto-metacognition-1993,th-Fan-2010-Beacons-Tasks} and may lack reliability~\cite{huang2019distilling, fry-patch-maintainability}. 

In contrast, a handful of recent research points to a link between the level of trust among users and how they visually scan information. Using eye--tracking technology, non--subjectively, and non--intrusively, they  shed light on the mental processes and perception of trust. Through capturing the dynamic patterns of visual focus, eye tracking unveils indicators regarding the users' cognitive processes~\cite{p-just-theory-reading-eye-1980,p-sharafi-practical-2020}. It is also useful in gauging mental workload during  task execution~\cite{p-sharafi-metrics-2015,p-sharafi-practical-2020,p-Poole-eyetracking-2005}, and in discerning the strategies and timings employed by participants while selecting and processing information~\cite{p-criticalLook-Goldberg-2010,p-just-theory-reading-eye-1980}.

A handful of previous studies have shed light on the relationship of trust in computer code. Alarcon \etal{}~\cite{p-alarcon-reputation-trust-2020} studied the impact of code authors' performance and reputation on developers' trust and their decisions to utilize the provided code. Several studies have explored the psychological underpinnings of trust in the context of automated program repair methods (cf.~\cite{alarcon-hsm-computer-code-trust-2018,ryan2019trust,p-alarcon-heuristic-systematic-2017,repairnator,fry-patch-maintainability,p-bertram-trust-eye-2020,p-Noller-trust-enhancement-2022}). Our work is closest in spirit to work by Bertram \etal{}~\cite{p-bertram-trust-eye-2020} that evaluated the perceived trustworthiness and potential biases developers might hold towards automated tools during code review. However, our focus is different. 

To the best of our knowledge, this paper is the first experimentally--controlled study via eye tracking to explore how participants' subjective trust and potentially biased actions are influenced by urgency and reputation. Regarding urgency, we alter the code while controlling for quality, labeling its priority as low or high. Also, we instantiate reputation by labeling the author's experience as senior vs. junior. 
Our 37 participants worked on six code review tasks evaluating Java code patches. We measure developers' tendency to trust and reuse the codes and performance coupled with visual attention trends and cognitive load~\cite{p-sharafi-practical-2020,p-sharafi-objective-2020} to provide insights into developers' cognitive processes when performing tasks.  
 

\vspace{1px}
We observe significant differences in the patch evaluation by all participants, depending on the perceived priority of the patch:

\begin{itemize}
    \item The patch priority had a statistically significant effect on task time ($p = 0.04$). Also, patches labeled as high--priority demanded significantly more visual effort ($p = 0.03$) --- indicative of cognitive load --- during evaluation.
   \item Different attention distributions in eye movements was observed when participants reviewed patches classified as low vs. high priority ($p = 0.001$).
    \item Participants did not self--report any influence of patch priority on their decision to accept, deploy, and reuse the code. Yet, they rated high--priority patches as higher quality, even when we controlled for quality. 
\end{itemize}

Also, we notice some disparities in the way all participants assess code patches, influenced by the apparent experience of the author:

\begin{itemize}
    \item Our findings revealed no significant differences in participants' performance and outcomes based on the author’s experience level.
    \item Eye--tracking data indicated varying attention allocation patterns and viewing strategies while reviewing patches authored by novice vs. senior authors.
    \item Our post--survey data did not indicate any effect of the author's experience on their assessments to trust and reuse.
\end{itemize}

Behaviorally, the thematic analysis of participants' self--report data shows that code functionality, quality, and comprehensibility are the three most important factors for code evaluation. 

Despite evident changes in their code review behavior in terms of associated visual and cognitive processes, our participants surprisingly failed to recognize the significant impact of urgency and reputation on their decisions to review, incorporate, and reuse the code. This  observed lack of awareness or understanding regarding the vital factors influencing their decision--making process during code reviews underscores the crucial significance of this research. Although code reuse is a common and profitable practice in the tech industry, an over--reliance or miscalibration of trust can lead to overuse and ignoring the constraints/potential dangers~\cite{alarcon-hsm-computer-code-trust-2018,p-alarcon-reputation-trust-2020,p-bertram-trust-eye-2020}. Consider OpenSSL, a common cryptography library. In 2015, the `Heartbleed' vulnerability let hackers access protected data on 24--55\% of popular websites, causing substantial reputational and financial harm~\cite{p-Durumeric-Heartbleed-2014}.

This novel paper expands the existing body of literature on the influence of trust on developers' performance and behavior. The results hold potential implications for developing training programs,  decision--making guidelines, collaborative review practices, and code review tools that provide feedback and raise awareness about the important cues towards creating a more objective and informed code review culture.

\section{Related work}

Related work includes studies of trust in computing~\cite{p-lee-trust-automation-2004,p-alarcon-reputation-trust-2020,alarcon-hsm-computer-code-trust-2018,ryan2019trust,p-Noller-trust-enhancement-2022,p-bertram-trust-eye-2020,kim2013automatic} and human studies that utilized eye tracking to examine code review tasks~\cite{p-crosby-algorithm-1990,p-uwano-eye-code-review-2006,p-sharif-eye-code-review-2012,p-begel-eye-coder-eview-2018,ford2019beyond}.

Concerning eye tracking and code review, our work takes the next step towards a better understanding of developers' code--reviewing behaviors and the various elements that shape their judgment and decisions. 

In terms of trust in computing, this paper is the first empirical study that contributes a fresh perspective on how developers perceive the trustworthiness of controlled quality patches based on urgency and reputation (\ie{} their apparent priority level and the author's experience). Moreover, most of the previous work used static stimuli, comprising small code snippets, while we covered indicative scenarios within an IDE (admitting editing, scrolling, and code  and unit--tests execution) in which participants worked at their own pace on a large code base.

\subsection{Trust in Computing} 


There is a rich body of work on how developers perceive and understand source codes~\cite{p-Rodeghero-automayed-doc-2017}. Yet, only a handful of studies  investigated the trustworthiness of code from developers' perspective and the psychological processes behind the perception of trust~\cite{p-alarcon-reputation-trust-2020,alarcon-hsm-computer-code-trust-2018,p-bertram-trust-eye-2020,p-Noller-trust-enhancement-2022}. 

Several experimental findings highlight the utility of the heuristic-–systematic processing model in the context of the trustworthiness of computer code. These results show the role of readability and the source's reputation in the accurate evaluation and reliability of the code~\cite{alarcon-hsm-computer-code-trust-2018,p-alarcon-reputation-trust-2020}.

The focus of the research on trust in software engineering is mainly on the effect of automation on trust. Lee and See~\cite{p-lee-trust-automation-2004} argue that the lack of trust in automated methods leads to their dismissal by developers. Alarcon \etal{}~\cite{p-Alarcon-Repair-Trust-2020,ryan2019trust} performed user studies and analyzed developers' self--reported rate of the trustworthiness of patches. They reported a determining impact of the patch's source on developers' behavior and indicated a higher degree of trust toward human--written patches. Fry \etal{} ~\cite{fry-patch-maintainability} studies the understandability and maintainability of machine-generated patches. Their findings suggest a potential disconnect between what human programmers perceive as crucial for program maintainability and the actual factors contributing to improved maintainability. Kim~\etal{}~\cite{kim2013automatic} developed candidate patches based on specific patterns and reported that developers were more inclined to accept these pattern-based patches. Bertram \etal{}~\cite{p-bertram-trust-eye-2020} carried out an eye--tracking study to explore how the source of a patch influences developers' code review behavior. They reported significant differences in code review behavior and found a preference for human-written patches, noting their readability and coding style. Noller~\etal{}~\cite{p-Noller-trust-enhancement-2022} conducted a survey involving over 100 developers to gain insight into the setup and parameters of artifacts and tools that could bolster trust in automated program repair. 


\subsection{Eye Tracking and Code Review} 

Previous studies have leveraged eye tracking to study how developers perform code reviews. They delved into developers' gaze patterns during code review and investigated how expertise can shape viewing strategies~\cite{p-crosby-algorithm-1990,p-uwano-eye-code-review-2006,p-sharif-eye-code-review-2012}, pinpointed potentially problematic code elements based on this viewing patterns~\cite {p-begel-eye-coder-eview-2018}, and explored the effect of additional technical indicators such as the number of followers and activity levels on the acceptance of patches~\cite{ford2019beyond}.
Huang \etal{}~\cite{p-Huang-Bias-2020} utilized medical imaging and eye tracking to investigate the neurological correlates of biases and differences between genders of humans and machines. They reported variances in code review behavior and biases based on the perceived author.

\section{Experimental Methodology}
We recruited 37 participants for this exploratory study investigating how developers interact with and use AI assistance while working within a large open-source project. Each participant was assigned two bug-fixing tasks of equal difficulty and collaborated with either a peer or GitHub Copilot as their source of assistance.

\subsection{Experiment Factors}

We evaluated the effects of two independent variables in our study: urgency, represented by priority level, and a trust cue, indicated by the author's experience, on developers' performance and code review behavior. 

\vspace{1px}
\textbf{The priority level:} Priority was presented as either ``High'' or ``Low''. Issue tracking systems such as Jira usually propose 5 priority levels: lowest, low, medium, high, and highest. Nevertheless, previous research has indicated that priority levels in the project are not well-defined and are inconsistently utilized~\cite{p-Baysal-non-technical-2013,p-Kononenko-codereview-quality-2015,p-Herraiz-simplify-bug-report-2008}. Thus, we decided to offer two levels to distinct urgent cases from the others.

\vspace{1px}
\textbf{The author's experience:} Experience in our study consisted of two categories: ``Senior'' and ``Junior.'' Following the pattern observed in online freelance marketplaces, we included the author's title and corresponding salary. In the bug report, the senior author was referred to as J. Miller, with the title and salary specified as ``Senior Developer -- \$100/hr.'' On the other hand, the junior author was referred to as M. Smith, identified as a ``Developer -- \$40/hr.'' 

To avoid bias from various human factors~\cite{ford2019beyond,p-Huang-Bias-2020}, including gender, age, attractiveness, and emotional facial expressions, we refrained from using the profile pictures and utilized initials instead of the participants' full first names. 

\subsection{Experiment Measures}
We assessed both the performance of participants during code review tasks (\ie{} objective behavior during code review) and their trust and reuse intentions (\ie{} subjective self--evaluations).

\vspace{1px}
\textbf{Performance}: To measure participant performance, first, we measured the amount of cognitive load or visual effort using eye--tracking data, which helped us understand the participant's level of engagement. Second, we tracked the total time spent by participants to complete the tasks, giving us insights into task efficiency. Finally, we recorded the number of correct answers provided by participants, indicating their accuracy in reviewing the code.

Eye gaze data was categorized into fixations and saccades~\cite{r-EMiRaIP-R78}. \textit{Fixations} are stable eye gazes lasting around 200--300 milliseconds. Researchers in psychology affirm that fixations are crucial for information acquisition and processing~\cite{p-just-theory-reading-eye-1980,p-Poole-eyetracking-2005}. \textit{Saccades}, on the other hand, are rapid eye movements that occur between fixations and involve minimal cognitive processing. We also analyzed eye gaze data based on specific \emph{areas of interest} (AOIs)~\cite{p-criticalLook-Goldberg-2010,p-sharafi-practical-2020} defined in the stimulus, such as 1) bug report text, 2) entire class files with patched code, 3) specific methods with patched code, 4) entire unit test files with patch evaluation, 5) and specific unit tests evaluating the patch. We used three standard eye--tracking metrics to measure the cognitive load: fixation count, average fixation 
duration and total fixation time~\cite{p-sharafi-metrics-2015,p-Poole-eyetracking-2005}. 
Fixation count and total fixation time indicate how visual attention is dispersed and how efficiently relevant data is found. Average fixation duration reflects the concentration of attention at specific areas. Higher values may imply information extraction difficulties and working memory strain ~\cite{p-criticalLook-Goldberg-2010,p-sharafi-practical-2020}. 

\textbf{Reuse intention}: Participants specified whether they reuse the patch by accepting and merging it into the code base to be passed over to the end user.

\vspace{1px}
\textbf{Trust intentions}: In addition to performance measures, to capture the nuanced responses and provide a quantitative measurement, we used six Likert scale questions for quality comparison of various patches:

\begin{enumerate}[leftmargin=5mm]
    \item How do you think about the coding style of the patch? (Very bad (1) -- Excellent(5)) 
    \item How do you think about the readability of the patch? (Very bad (1) -- Excellent(5))  
    \item How do you think about the comment (summary) the author provided for the patch? (Not clear at all (1) -- Very clear (5)) 
    \item How much do you think the patch deploys the functionality it claims in the summary? (Very little (1) -- Very much (5)) 
    \item How do you rank the general quality of the patch? (Very bad (1) -- Excellent(5))
    \item How trustworthy do you find this code? (Untrustworthy (1) -- Trustworthy (5)) 
\end{enumerate}

\subsection{Participants and Recruitment}
In our study, which was approved by our Institutional Review Board (IRB), we recruited a total of 39 participants, comprising undergraduate and graduate students from the authors' institutions. Among these participants, two were recruited for a pilot phase to ensure the smooth setup of the experiment and verify its validity. The data obtained from the other 37 participants were used for our study analysis and findings.
Participants completed questionnaires to collect essential demographic information, and the summarized details can be found in Table~\ref{tab:demographics}. Among all the recruited students, we inquired about their gender, including options for male, female, and non-binary. However, no participant identified themselves as non-binary. We reached out to participants via email and offered them a \$25 voucher as compensation for their participation.

We also inquired about participants' experience and familiarity with programming, Java, and working with Integrated Development Environments (IDEs). On average, participants reported six years of programming experience (SD = 2.92). All participants were well--versed in Object--Oriented Programming (OOP), had a good understanding of Java, and possessed prior experience working with IDEs. We relied on participants' self--estimation of their programming experience, a method proven to be reliable when working with students, as evidenced by Siegmund \etal{}~\cite{p-siegmund-measuring-expertise-2014}.

\subsection{Software System and Task}
In this experiment, we selected \emph{jFreeChart} as the system under study. \emph{jFreeChart} is a Java--based open--source project that allows users to create and display charts within their applications. Specifically, we utilized version 1.1.0, which was released in 2015 and comprised approximately 300 KLOC (thousand lines of code) and around 94,000 Java classes. To ensure the authenticity of our code review tasks, we handpicked six historical bugs from a predefined set that marked as resolved, as listed in Table~\ref{tab:tasks}. These bugs were chosen to represent realistic scenarios.

In our evaluation process, we included a variety of patches with (experimentally-controlled) different labels to represent the (purported) expertise level of the authors (Senior or Junior) and the (purported) urgency of the patches (high priority or low priority). These patches were chosen from Defects4j-Repair project~\cite{martinez2016}. Each patch consisted of a code sample, ranging in size from 8 to 76 lines of code (Mean: 11, SD: 13), with repaired lines varying from 3 to 41 (Mean: 26, SD: 23). We randomly ordered the six code review tasks for each participant to ensure fairness and confirm that all participants reviewed all six patches, considering all possible combinations of different authors and priorities. 

\begin{figure}[t]
\centering
\frame{
\includegraphics[scale=0.35]{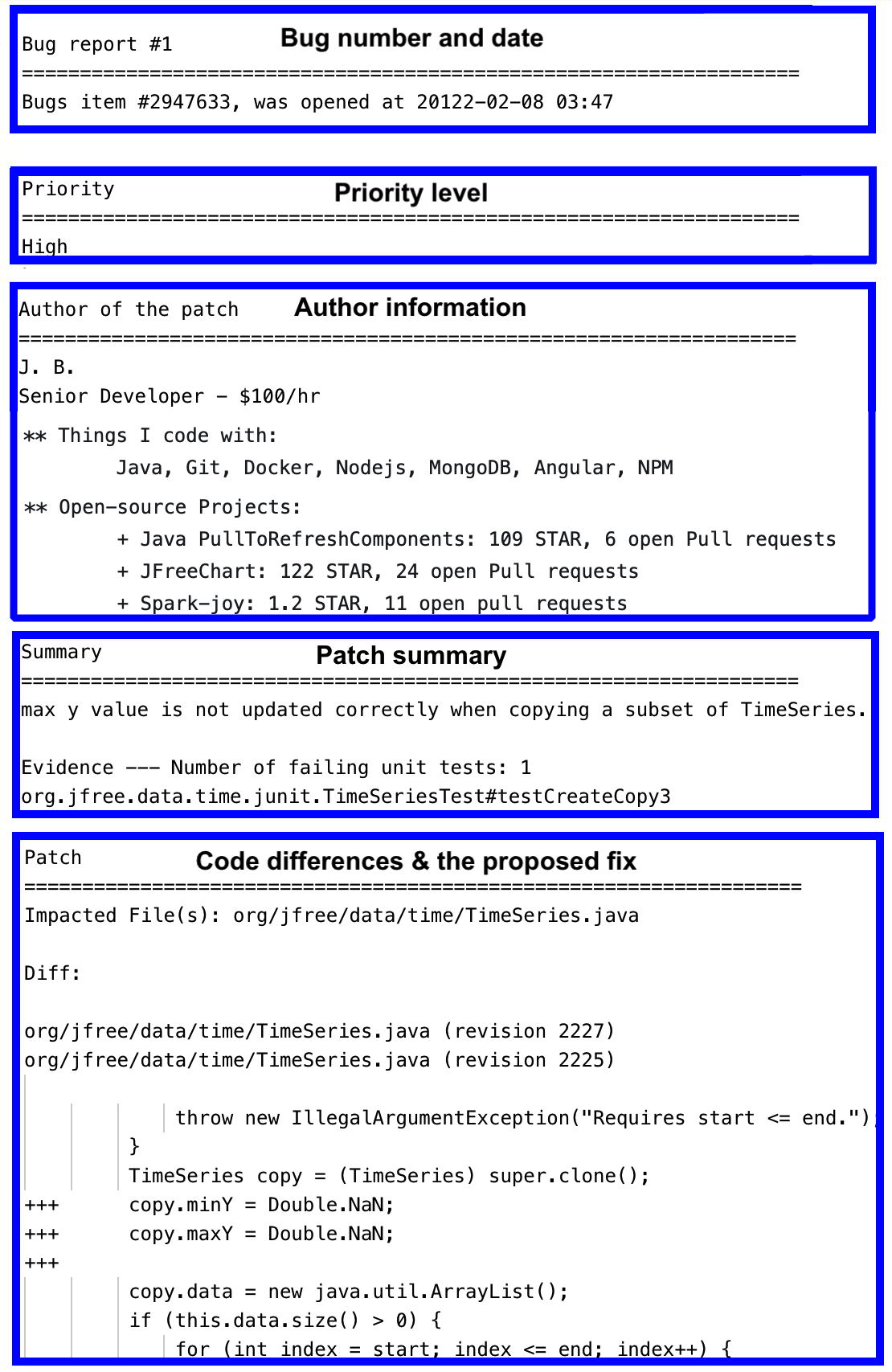}}
\caption{Distribution and intensity of visual attention across the Code Pane and Stack Trace AOIs for two assistance types (Copilot and Peer), averaged across all participants. Participants exhibited greater visual effort in evaluating outputs and errors when using Copilot, and also showed increased focus on the Project Structure pane.}
\label{fig:bug-report}
\end{figure}

Each code review task involved examining a bug report file, which contained the following information: 1) the bug report number and the date of report, 2) the (controlled) author's information, including their name,  their salary, and skills, 3) the (controlled) patch's priority level (high or low), 4) the summary of the bug, 5) the number and names of failing unit tests, and 6) the actual patch itself, demonstrating the code changes made to fix the issue. An example of a bug report file is shown in Figure~\ref{fig:bug-report} with various parts highlighted. Throughout the experiment, participants had unlimited access to the entire code repository and could freely navigate it using the Eclipse IDE.

\begin{table}[t]
  \caption{Participant demographics: age, gender, \& study level.}
  \label{tab:demographics}
  \begin{tabular}{lccc}
                    & All  & Men  & Women  \\ 
    Characteristics & \textit{(N = 37)} & \textit{(n = 25)} & \textit{(n = 12)}  \\ \hline 
    Age (n (\%))    &                         \\
    \rule{1ex}{0pt} 18-25  &      18 (48\%)  & 16 & 2  \\
    \rule{1ex}{0pt} 25-30  &     10 (27\%)   & 5 & 5  \\
    \rule{1ex}{0pt} 30-35  &      5 (13\%)  & 2 & 3  \\
    \rule{1ex}{0pt} 35 or older  &      4 (10\%)  & 2 & 2  \\
    \rule{0pt}{3ex}  Study level (n (\%))    &                      \\
    \rule{1ex}{0pt}  First year &  2 (5\%) & 2 & 0      \\
    \rule{1ex}{0pt} Sophomore & 2 (5\%)   & 1 & 1\\
    \rule{1ex}{0pt} Junior & 4 (10\%)   & 3 & 1\\
    \rule{1ex}{0pt} Senior & 1 (2\%)   & 1 & 0\\
    \rule{1ex}{0pt} M.S./PhD & 28 (75\%) & 18 & 10  \\
    \midrule
\end{tabular}
\end{table}

\begin{table*}
\caption{Explanation of the patches, including a brief summary of the reported bug alongside the affected sections of code.}
\label{tab:tasks}
\begin{tabular}{p{120mm}p{15mm}p{17mm}p{12mm}}
\hline
&  \multicolumn{3}{c}{\textbf{Scope of the solution}} \\ 
\textbf{Short Description} & Classes (Methods) & Test Classes (Unit Tests) & Lines of code\\\toprule
Patch 1: Max-y value is not updated when copying a subset of TimeSeries.    & 1 (1) & 1 (1) & 9\\ \hline
Patch 2: \textit{XYSeries.addOrUpdate()} should handle duplicate X values, as a prior change supported. Yet, the method was not updated, resulting in overwriting of the data.
 & 1 (1) & 1 (1) & 14\\ \hline
 Patch 3: Potential Null Pointer Exception in \textit{AbstractCategoryItemRender.getLegendItems()} results in a null pointer access in the class.  & 
1 (1) & 1 (1) & 10\\ \hline

 Patch 4: If the label generator returns null, the PieChart must be created. & 1 (1) & 1 (1) & 27\\ \hline

 Patch 5: The cached bounds must be reset whenever items are added or removed from the plots. Now, we need to iterate over the entire dataset to update the cached values. &
1 (4) & 1 (1) & 76\\ \hline

Patch 6: We have a null pointer exception in \textit{StatisticalBarRenderer} when one of a series in a category has no data. 
& 1 (1) & 1 (4) & 24 \\ \hline

\end{tabular}
\end{table*}

\subsection{Equipment and Setup}
For our experiments, we conducted them using a 27" monitor with a screen resolution of 1920x1080 pixels. To track eye movements, we utilized the Tobii Pro Fusion eye-tracker \cite{s-tobii}. The Tobii Pro X3--120 is a remote, non-intrusive device that captures gaze data at 250 Hz and has the capability to precisely locate eye-gaze data within a code document, down to the granularity of a single line of 10pt text. 

To provide a realistic environment with indicative code review tasks, we facilitated tasks such as scrolling, switching between files, and editing code by installing and utilizing the iTrace plugin~\cite{shaffer2015itrace}. This state--of--the--art plugin allowed participants to interact with the source code and 
other relevant materials naturally while also collecting the necessary measurements. Using iTrace Toolkit, the recorded raw data was then subjected to the Velocity Threshold Identification (I--VT) filter to generate fixations.

\subsection{Procedure}
We conducted the experiment in a quiet room with an eye tracker. Participants were seated approximately 70 cm from the screen in a comfortable swivel chair with armrests. Before running the experiment, all participants signed a consent form, and the experimenter verbally explained the experiment’s procedure in detail. Participants were informed that the experiment consisted of one Java project and that they would be sequentially reviewing six patches. The experimenter did not explain the particular goal of the experiment. Participants were given ten minutes per code review task and were instructed to inform the experimenter if they finished early to stop the tracking. To mitigate any learning effects, participants received the six tasks in randomized order. To better control and avoid any other factors impacting the results, the participants were instructed to maintain the full--screen IDE setup, not use the debugger, and not browse the internet.

Upon completing each task, participants were presented with a survey that encompassed various aspects. The survey included questions regarding the overall quality of the patches, the level of trust for critical tasks, the quality of implemented functionality, patch readability, and patch coding style. Participants were also required to indicate their preference of ``accept'' or ``reject'' for the corresponding task within the survey. Following the completion of all six code review tasks and their associated surveys, participants were asked to complete a post-questionnaire.

To decrease potential stereotype threat~\cite{p-shapiro2007stereotype}, we strategically placed questions about coding knowledge and experience at the end of the study. Specifically, women and marginalized groups encounter the detrimental perception that their abilities are inferior more profoundly than individuals from other backgrounds~\cite{p-steele1995stereotype}. Additionally, the post-questionnaire contained inquiries about participants' age range, gender, population group, native language, level of study, field of study, years of programming experience, the influence of the author's expertise level, and the influence of urgency in accepting or rejecting the patches.

\section{Data Analysis and Results} 

This study investigated how developers perceive the trustworthiness of code patches, the extent to which the perceived priority level of the patches, and the experience level of the author affect developers' efficiency and behavior during code review. We ensured the code remained consistent while we varied the patch's listed priority level (high or low) and the author's listed experience (senior or junior). Our analysis primarily centered on the responses to the following research questions:

\begin{itemize}
    \item[\bf RQ1.] How well do participants' self--reports regarding the role of patches' priority level and author's experience correspond with our collected data?
  \item[\bf RQ2.] What effect do the priority levels of patches and the author's experience  have on participants' performance in code review?
    \item[\bf RQ3.] How are the participants' code review behaviors influenced by the priority levels of patches and the author's experience during code review? 
\end{itemize}

We provide access to our study material and de--identified dataset, consisting of behavioral data, eye--tracking data, and survey data, for analysis and replication at link.

\begin{figure*}[t]
\centering
\includegraphics[scale=0.40]{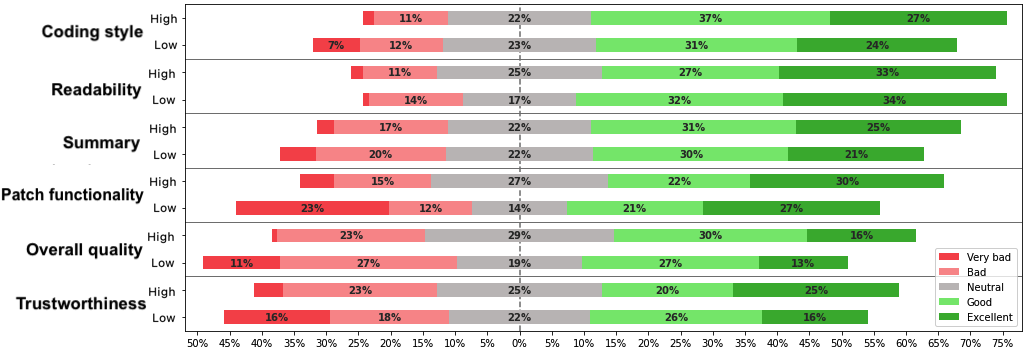}
\caption{Ratings of patch quality attributes and trustworthiness by the priority level with a 5--point Likert scale (222 responses). ``High''  priority patches were rated higher in coding style, functionality, trustworthiness, and in a statistically significant manner $(p=.03)$, superior in overall quality by participants.}
\label{fig:rating-priority}
\end{figure*}

\begin{figure*}[t]
\centering
\includegraphics[scale=0.40]{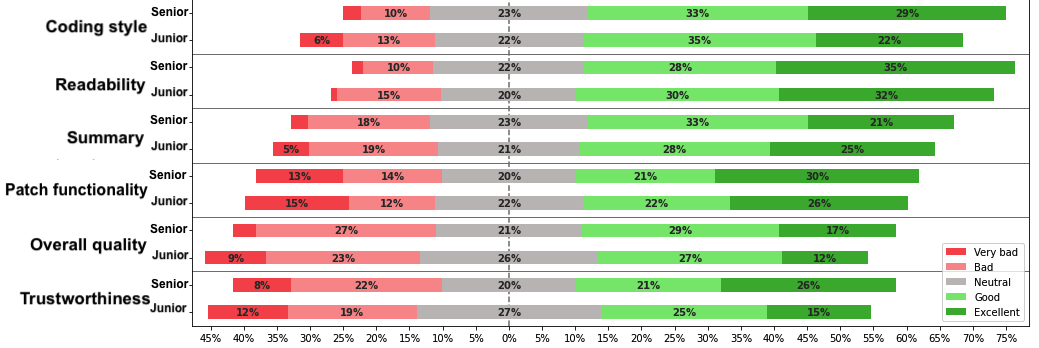}
\caption{Ratings of patch quality attributes and trustworthiness by the author's experience with a 5-point Likert scale (222 responses). Patches labeled as written by a senior developer were generally ranked higher.}
\label{fig:rating-experience}
\end{figure*}

\subsection{RQ1. Self--Reporting and Trust}
\label{sec:RQ1}

In our study, all 37 participants answered the post--questionnaire regarding the tasks and their experiences to provide insights into their subjective evaluations of trust.

The participants answered six questions concerning the overall quality of the patches, the level of trust for critical tasks, the quality of implemented functionality, patch readability, summary of the patch, and patch coding style. The results of the Likert scale questions for the priority level and author's experience are presented in Figure~\ref{fig:rating-priority} and Figure~\ref{fig:rating-experience}, respectively. Our data reveals interesting patterns in how patch priority levels are perceived. 

The participants tended to assign higher rankings to patches labeled as  ``High priority'' for the overall quality, and the result is statistically significant (Mann-Whitney test, $W = 6910.5, p = .03$). The proportion test (Chi-squared test for significance) regarding patch acceptance and the Wilcoxon signed-rank test for trust levels did not distinguish between high and low--priority patches. Participants, on average, were 6\% less likely to accept patches classified as low--priority. 68\% of participants indicated that the perceived urgency or severity of the issues addressed by the patches did not impact their decision to accept or reject them. Yet, high--priority patches were rated as higher quality by participants. This finding suggests that even if the issue's urgency does not directly affect their acceptance decision, it does affect their perception of the patch's quality.

In addition, we found no evidence that the author's experience impacted participants' self--reported data related to quality factors. Around 60\% of the participants mentioned that the experience level of the authors doesn't influence their judgment on whether to accept and or reject the patch. We found no compelling evidence suggesting that the author's experience level significantly impacts participants' intention to reuse.

\begin{tcolorbox}[colback=bubbles!5!white,colframe=bubbles!75!black,left=0pt, right=0pt,top=1pt,bottom=0pt,title=RQ1 --- Likert scale self-reporting and trust]

\textbf{Patch priority:} Participants did not indicate any impact of the patch's priority on their evaluations and intention to reuse. High--priority patches were rated higher in overall quality, despite controlling for quality.
  \vspace{-5pt}
  \tcblower
  \vspace{-5pt}
  \textbf{Author's experience:} There was no evidence to support the author's experience affecting participants' self-reported quality factors.
\end{tcolorbox}

To avoid biasing participants' self-reports in a specific direction, we also utilized open--ended free response questions:

\begin{enumerate}[leftmargin=5mm]
    \item Regarding prioritization, what were the top three key factors you considered when making decisions during code reviews?
    \item Does the urgency and severity of the issue affect your decision to accept or reject the patch? Please clarify.
    \item Does the expertise of the patch's author affect your decision to accept or reject the patch? 
\end{enumerate}

\begin{figure}[t]
\centering
\includegraphics[scale=0.28]{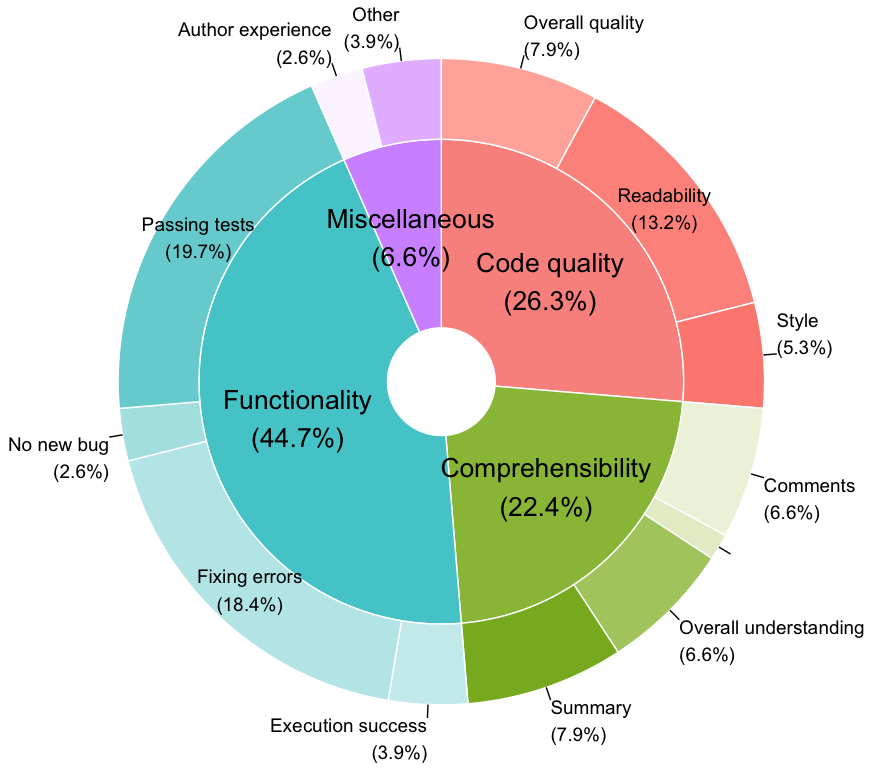}
\caption{A Breakdown of the top three self-reported factors and their subcategories in code review evaluation, displayed with self-reported frequencies. Three key factors that guide code review decisions are functionality (highlighted by 44.7\% of participants), code quality (26.3\%), and comprehensibility, mainly represented by the quality of the comments and summaries (22.4\%). }
\label{fig:treeplot}
\end{figure}

We carried out a thematic analysis of responses from all 37 participants to the aforementioned questions, focusing on the crucial factors they used to assess code patches. Through a two-step process involving descriptive coding and subsequent discussions, we identified and grouped these essential elements. As shown in Figure~\ref{fig:treeplot}, the findings revealed three primary factors that participants considered: 1) functionality and whether the code passes tests was the most significant factor, with 44.7\% of participants mentioning that whether the code functions as expected and passes all tests is crucial to their decision-making process, 2) code quality with a focus on readability account for 26.3\% of responses and was the second most influential factor, and 3) code comprehensibility and the quality of comments and summaries (22.4\%). Other factors accounted for the remaining 6.6\% of responses. Notably, only 2.6\% of the participants reported the author's experience as a critical factor in their decision-making process. The self--reported data show that in the context of code review, the merit of the code itself outweighs considerations about the author's experience or the patch's priority. 

\begin{tcolorbox}[colback=bubbles!5!white,colframe=bubbles!75!black,left=0pt, right=0pt,top=1pt,bottom=0pt,title=RQ1 --- Qualitative analysis of self-reporting and trust]
    Functionality, quality (readability and style), and code comprehensibility are the top three factors in patch evaluation.
\end{tcolorbox}

\begin{table}[t]
\centering
\caption{Pairwise comparisons of performance metrics using the Chi--squared test for accuracy and acceptance rate and non--parametric Wilcoxon Test ($\alpha = 0.05$) for time and visual effort metrics for priority levels (High vs. Low). Significant results ($p <0.05$) are bolded. A significant trend showed that participants devoted more visual attention and task time to analyzing high--priority patches.}
\label{tab:performance-priority}
\begin{adjustbox}{width=\columnwidth,center}
\begin{tabular}{lccl} \toprule
\hline
    & \multicolumn{2}{c}{Mean (Std. Dev.)}   \\ 
    \textbf{Priority level} & High           & Low &  $p$ \\ \midrule
\multicolumn{1}{l|}{Acceptance rate} & 56\% & 50\% & 0.5 \\
\multicolumn{1}{l|}{Accuracy}   &  70\%      &   63\%      & 0.3 \\
\multicolumn{1}{l|}{Time spent (s)} & 449 (161) & 406 (153) & \textbf{0.04}  \\
\multicolumn{1}{l|}{Avg. Fx. Duration (ms)} & 647.8 (211.7)  & 644.8 (209.2)  & 0.7 \\
\multicolumn{1}{l|}{Fx. Count} & 428.9 (190.9) & 398.3 (61.2) & 0.1 \\
\multicolumn{1}{l|}{Total Fx. Time (s)} & 271 (123)  & 236 (104) & \textbf{0.03} \\\hline  
\end{tabular}
\end{adjustbox}
\end{table}

\begin{table}[t]
\centering
\caption{Pairwise comparisons of performance metrics using Chi--squared test for accuracy and acceptance rate and  non--parametric Wilcoxon Test ($\alpha = 0.05$) for time and visual effort metrics for experience level (Senior vs. Junior). Significant results ($p <0.05$) are bolded. We found no impact of the the author's experience on participants' performance and acceptance rate.}
\label{tab:performance-experience}
\begin{adjustbox}{width=\columnwidth,center}
\begin{tabular}{lccl} \toprule
\hline
    & \multicolumn{2}{c}{Mean (Std. Dev.)}   \\ 
    \textbf{Author's experience} & Senior   & Junior &  $p$  \\ \midrule
\multicolumn{1}{l|}{Acceptance rate}  &  54\%  & 57\% & 0.9 \\
\multicolumn{1}{l|}{Accuracy} & 68\% & 65\%  & 0.6\\
\multicolumn{1}{l|}{Time spent (s)}  & 429 (150) & 420 (159) & 0.9 \\
\multicolumn{1}{l|}{Avg. Fx. Duration (ms)} & 640.4 (165.5) & 652.6 (249.2)  & 0.8  \\
\multicolumn{1}{l|}{Fx. Count} & (168.5) & 408.0 (186.6) & 0.5 \\
\multicolumn{1}{l|}{Total Fx. Time (s)} & 257 (111) & 252 (120) & 0.8 \\\hline  
\end{tabular}
\end{adjustbox}
\end{table}

\subsection{RQ2. Performance Differences}


We investigated whether the patches' priority level and the author's experience impact the participants' overall performance measured by accuracy, total time spent, and visual effort. The results, as summarized in Table~\ref{tab:performance-priority} and Table~\ref{tab:performance-experience}, were analyzed using proportion tests (Chi-squared test) for our categorical variables, accuracy (0 \slash 1) and acceptance rate (Accept\slash Reject) and Wilcoxon signed-rank test for total time as a continuous variable. 

To assess visual effort, we examined fixation count (Fx. Count), average fixation duration (Avg. Fx. Duration), and total fixation time (Fx. Time), indicators of cognitive load. A higher number of fixations and longer fixation duration suggest increased visual effort and mental load or denote the importance of the focused area. Previous studies have reported the influence of participants' trust within the system on various eye-tracking metrics~\cite{lu-sarter-eye-trust,p-gold-trust-2015,p-geitner-trust-tech-2017}. 

The findings revealed no significant differences in acceptance rate, accuracy, fixation count, and average fixation duration based on either the patches' priority level or the author's experience level. However, as shown in Table~\ref{tab:performance-priority}, we observe longer fixation time and task time for high--priority patches in a statistically significant manner. 
The perceived urgency or potential impact associated with high--priority patches led to a more meticulous evaluation of code patches by our participants.

\begin{tcolorbox}[colback=bubbles!5!white,colframe=bubbles!75!black,left=0pt, right=0pt,top=1pt,bottom=0pt,title=RQ2 --- Performance differences]

\textbf{Patch priority:} Patches labeled as high--priority
demanded significantly more task time and visual effort (\ie{} cognitive load).

  \vspace{-5pt}
  \tcblower
  \vspace{-5pt}
\textbf{Author's experience:} No significant performance metric variations are tied to the author's experience level.
\end{tcolorbox}

To gain further insights, we analyzed how equal amounts of time were allocated differently based on urgency and reputation.

\subsection{RQ3. Differences in Code Review Behaviour}

We examined how participants' code review behavior and problem--solving strategy are influenced by the priority level of patches and the authors' experience with an analysis of attention distribution and navigation trends over time throughout a task. We calculated our three eye--tracking metrics (average fixation duration, fixation count, and total fixation time) on each AOI and compare the distribution of attention across AOIs for patches labeled as high vs. low--priority and senior vs. junior--authored.

\begin{figure}[t]
\centering
\includegraphics[width=1.0\columnwidth]{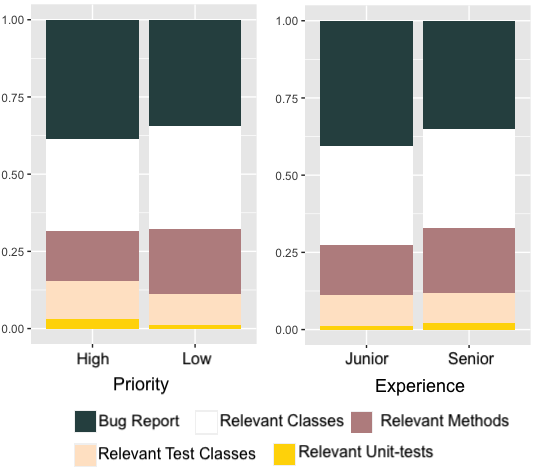}
\caption{Distribution and the intensity of visual attention across AOIs for various priority and experience levels, averaged for all participants. The participants spent more visual effort evaluating the relevant unit tests for high--priority patches. In contrast, they spent more effort on relevant methods for low--priority patches. Also, participants focused more on relevant methods for senior labeled patches. }
\label{fig:dist}
\end{figure}

For priority level, the general align-and--rank non--parametric factorial analysis, as described by Wobbrock \etal{}~\cite{p-wobbrock-ART-2011}, indicates a significant interaction between the marked priority level and the distribution of visual attention for all three metrics (Fx. Count: $F(1, 5) = 3.6, p<.05$, Avg. Fx. Duration: $F(1, 5) = 2.7, p<.05$, and Total Fx. Time: $F(1, 5) = 4.1, p=.001$).

Our results also highlighted the significant interaction between marked author's experience and the visual attention distribution for two metrics out of three: fixation count, and total fixation time (Fx. Count: $F(1, 5) = 2.5, p<.05$, Avg. Fx. Duration: $F(1, 5) = 0.11, p=.9$, and Total Fx. Time: $F(1, 5) = 3.0, p=.009$). 

Figures~\ref{fig:dist} illustrates the distinct code scanning behavior of participants across different areas of interest (AOIs) when patches with various labels (low vs. high--priority and senior vs. junior--authored). These findings provide evidence that the perceived priority of patches and the author's experience significantly influence participants' perception of the relevance of different code sections. 

While higher--level performance metrics may not detect a noticeable difference, our analysis reveals that participants exhibit diverse scanning behaviors characterized by attention distribution and intensity variations during code review. Specifically, for priority level, we observed that participants exerted more significant visual effort when evaluating the relevant unit tests for high-priority patches. Conversely, participants invested more effort in examining relevant methods when reviewing low--priority patches. For the author's experience, we also observe more focused visual attention on relevant methods for senior labeled patches.

\begin{figure*}[t]
\centering
\includegraphics[scale=0.57]{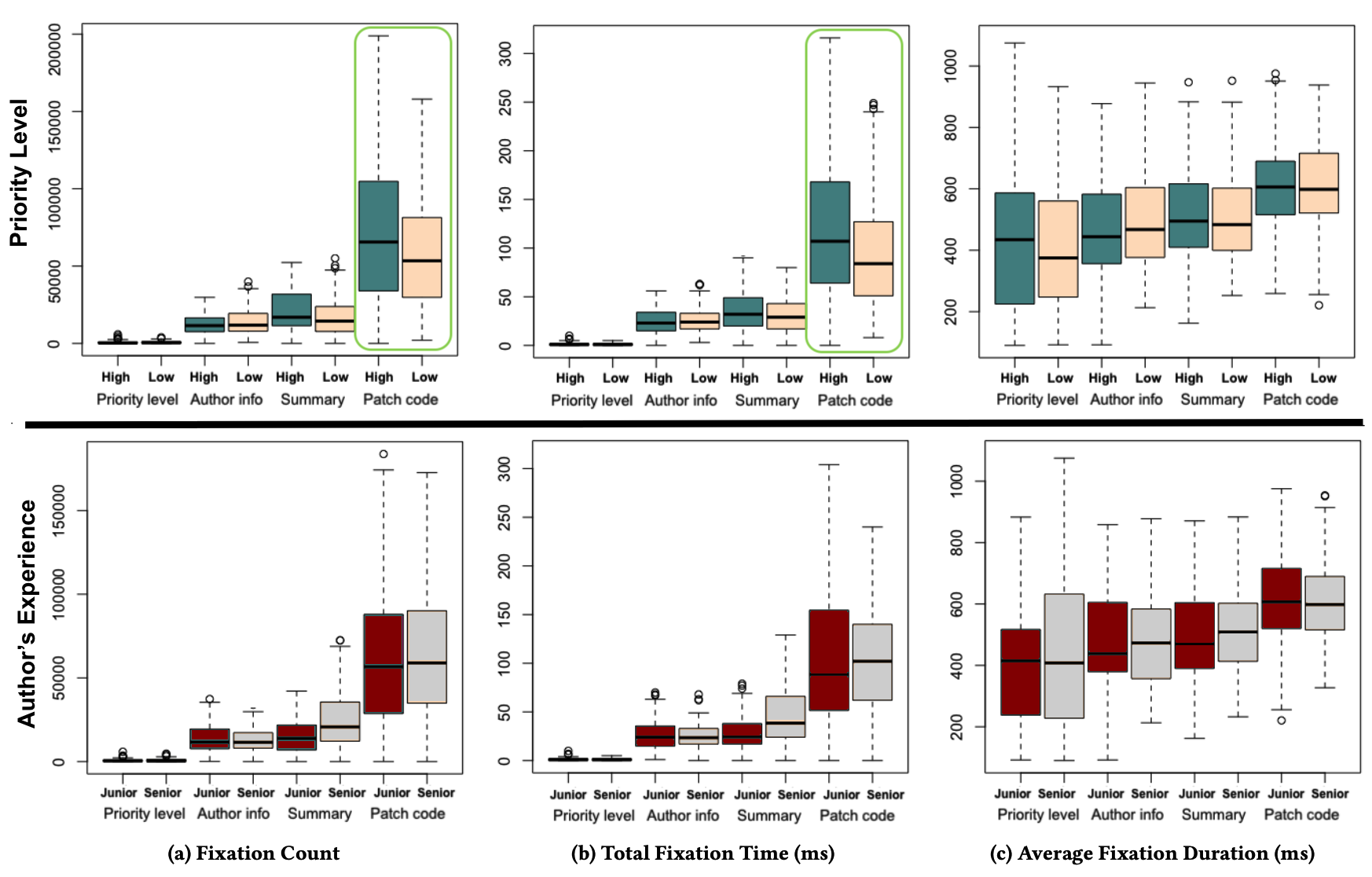}
\caption{Summary of visual attention distribution for various parts of the bug report files for all participants for priority levels (top row) and author's experience (bottom row). No significant effect of author's experience was found on eye--tracking high--level metrics in isolation. However, participants, on average, spent more time and effort (fixation count and total fixation time) analyzing the code part of high--priority patches, highlighted with green rectangles.}
\label{fig:priority-box}
\end{figure*}

In addition, we investigate how our participants spend their time reviewing the bug report and where they look.  More specifically, developers consider what elements and whether this varies with the patch's priority level or the author's experience.  We identified that participants fixated
more on the patch summary and the proposed code changes. Similar
to the results of our qualitative analysis of participants' self--report
data (cf. Section~\ref{sec:RQ1}), most participants focused on code and technical parts. Also, in a statistically--significant manner, our participants spent more time and cognitive effort (Wilcoxon test with Bonferroni adjustment:  $p = 0.03$ for Fixation count and Fixation time) analyzing the code part of the high--priority patches as shown in Figure~\ref{fig:priority-box}.

The findings shed light on the intricacies and challenges involved in comprehending developers' perception of trust towards various tools and artifacts. It underscores the advantages of adopting a multi-modal approach in exploring this complex phenomenon. Although we might not have observed a direct impact on performance metrics, our eye-tracking results reveal noteworthy variations in the allocation of visual attention and cognitive resources based on the urgency and reputation under investigation. These insights provide valuable implications for understanding how developers interact with and rely on various parts of the source code.


\begin{tcolorbox}[colback=bubbles!5!white,colframe=bubbles!75!black,left=0pt, right=0pt,top=1pt,bottom=0pt,title=RQ3 --- Code review behavior]

\textbf{Patch priority:} AOI relevance --- high--level scanning patterns of participants --- varies significantly for high and low--priority patches. Participants put more attention and cognitive load into evaluating the relevant unit tests and the code part in bug reports for high--priority patches.
  \vspace{-5pt}
  \tcblower
  \vspace{-5pt}
\textbf{Author's experience:} The experience level of authors impacts the overall viewing strategies of participants.  More attention on reading and processing relevant methods for patches labeled as written by senior developers. No significant differences in AOI relevancy were found as a function of the author's experience with bug reports.
\end{tcolorbox}

\section{Discussion of the Results}


Based on the dual process model of persuasion~\cite{b-persuation-petty1986}, individuals interpret information via two paths: central and peripheral. Central route or systematic processing relies on logical reasoning and analytical skills. On the other hand, peripheral processing employs heuristics, stereotypes, and biases to lessen the cognitive burden during decision-making~\cite{p-sunstein-moral-2005}. Peripheral processing tends to occur more when decisions are emotionally charged, trusted, or time-limited. In this study, to examine trust, we thoughtfully employ cues of urgency and reputation to stimulate peripheral processing. This method enables us to investigate biases and inclinations towards priority levels and authors' experience precisely, setting our research apart as the first to do so.

\subsection{The Role of the Priority Level} 
The findings of our study indicate a variance in how developers execute code reviews (\ie{} noticeable variation in attention allocation), which correlates with the priority of the given issue. We support these findings through eye tracking and behavioral data. In addition, the patch priority has a statistically significant effect on the time taken to handle the tasks. 

Contrary to some previous studies~\cite{p-Baysal-non-technical-2013,p-Kononenko-codereview-quality-2015}, our findings reveal no significant difference in the acceptance rate based on the priority level of patches, adding nuance to the existing understanding. While these studies suggested that high--priority patches were more likely to be incorporated into the project's codebase through post--factum analysis of GitHub patches, they also highlighted a pertinent issue. Developers often display confusion over the meaning and implications of standard priority levels, ranging from P1 (low) to P5 (high). This indicates a potential gap in the shared understanding and interpretation of these priority scales across the community.

Adding further to this complexity, our results suggested that participants often overlooked patch priority when determining whether to accept and reuse the code. This divergence could stem from the code's quality, functionality, or readability, overshadowing the assigned priority level in a developer's decision--making process. Alternatively, the mismatch could arise from the need to clarify what each priority level implies.

Thus, our findings call for revising and enhancing software development guidelines and code review tools to emphasize the significance of patch priority levels and to present assessment criteria clearly. These enhancements enable developers to prioritize urgent patches, resulting in streamlined decision-making, improved code quality, and software development efficiency.

\subsection{The Role of Experience Level}

Surprisingly, the author's experience had no effect on participants' trustworthiness perceptions, reuse intentions, or performance. A couple of previous work in computing and trust has demonstrated the impact of code reputation on developers' trust and reuse perceptions and outcomes~\cite{p-alarcon-heuristic-systematic-2017,p-alarcon-reputation-trust-2020}. In our research, we chose to focus on the human elements of trust and controlled the code's reputation through the experience level of the author. This approach differs from past work, which established reputation by either modifying the metadata attributes of OSS systems (for example, total commits, likes, and star ratings)~\cite{p-alarcon-reputation-trust-2020} or labeling the source as either reputable or unknown~\cite{p-alarcon-heuristic-systematic-2017}. By removing the profile picture and using initials, we avoided biases based on gender, age, attractiveness, and emotional facial expressions. In addition, we provided a complete profile for both novice and senior authors, which implies trustworthiness and professionalism~\cite{ford2019beyond}. Our findings prompt reevaluating how the software engineering community perceives and utilizes reputation and what constitutes an effective reputation metric. The implications would make the experience level less of a barrier for new or less experienced developers to be accepted and evaluated based on their technical merits.

Although we found no differences in our performance metrics, our analysis of the distribution of visual attention and the intensity of visual processing reveal different AOI preferences based on the author's experience. Our results highlight that the apparent experience of the patch's author influences how developers judge the code. These findings broadly align with these previous studies in finding statistically significant differences in how developers carry out the task of code review~\cite{p-bertram-trust-eye-2020,p-Huang-Bias-2020}. Their results also highlight the impact of the patch's provenance (men vs. women vs. machine) on developers' viewing strategies.

The reported differences in scanning patterns based on the author's labeled experience, revealed by the eye-tracking data, are likely attributed to differences in training or feedback. Previous work also stated that in addition to technical aspects, social ones (usually related to the author and presented in their user profile) also are being considered by developers when deciding upon patch review and acceptance~\cite{ford2019beyond}. In the same vein, Alarcon \etal{}~\cite{p-alarcon-reputation-trust-2020} reported more involvement with the code for more reputable patches.

Prior studies in automation have found a relationship between age and inclination to trust~\cite{ryan2019trust,alarcon-hsm-computer-code-trust-2018}. Given that our sample group consisted of young students with minimal professional experience, this could account for the lack of observable impact on performance metrics and trust self-evaluation associated with the author's experience in our findings.


\section{Threats to validity}

\textbf{Internal validity:} To minimize the instrument bias, we employed a video--based eye--tracking system that does not require cumbersome goggles and permits participants to move their heads without affecting the camera's calibration. The camera calibration was carefully executed at the start of the study and adjusted between tasks to ensure accuracy. To prevent the treatments diffusion, participants were advised against discussing the experiment. To mitigate the Hawthorne effect, we clarified that the eye tracker only captures eye-related data, and our experimenter was seated discreetly at a distance from the participants to minimize their sense of being watched. To mitigate stereotype threat and avoid influencing performance~\cite{p-spencer1999stereotype,p-shapiro2007stereotype}, we saved self--assessment of coding skills for the study's end. 

\textbf{Construct Validity:} We reduced the risk of hypothesis guessing and apprehension by withholding the exact aims of the study from participants. However, we did ensure they were well-informed about the study's procedure, session count, task types, and the functioning of the eye tracker prior to the experiment.

\textbf{External validity:} All of our participants were students, with more than 85\% being graduate students with good programming expertise. Using students as participants is acceptable when evaluating techniques for the novice or non-expert software engineers, as they represent the future generation of software professionals~\cite{p-Guidelines-Kitchenham-2002}. 
The choice of the JFreeChart project as the subject system poses a validity threat due to its potential influence on the study. However, we mitigated this risk by selecting a reasonably large and complex open--source project in a popular programming language. We also chose the historic bugs from the project repository that accurately represent real--world software issues.

\textbf{Conclusion validity:} To mitigate this threat, we utilized established eye--tracking metrics, employed standard statistical methods for data analysis, and conducted eye--tracker calibration for each participant before each task.

\section{Conclusion}

Deriving insights from the psychological aspects of trust and compliance, we conducted a carefully controlled experiment with 37 participants using advanced eye--tracking technology to delve deeper into the factors influencing developers' trust--based cognitive processes. This novel research offers a holistic picture of how urgency and reputation impact developers' perceptions of trustworthiness and evaluation during code reviews. 

Our qualitative and quantitative study indicates that patch priority significantly impacted developers' review behavior and their perception of code quality without affecting reuse decisions. Interestingly, the author's experience level of the code patches altered participants' attention distribution but not review performance. In addition, the most commonly self--reported factors in code evaluation that affect our participants’ decisions were: (1) functionality,  (2) code quality and (3) code quality.


The rising trend of reusing auto--generated code, such as Facebook's SapFix~\cite{MargineanSapFix2019}, and collaborating with AI pair programmers like GitHub's Copilot~\footnote{https://github.com/features/copilot}, underscores the critical role of trust in software development. It also prompts exploring how much developers should rely on and reuse such generated code. Our research represents a significant step forward in understanding the dynamics of trust in software engineering tasks from developers' perspectives. By gaining a deeper understanding of factors underlying the trustworthiness of software artifacts, we can work towards enhancing the efficiency of code reuse practices and code automation, ultimately leading to more robust and dependable software systems.

\begin{acks}
To Robert, for the bagels and explaining CMYK and color spaces.
\end{acks}

\nocite{*}
\bibliographystyle{ACM-Reference-Format}
\bibliography{ref}

\appendix

\end{document}